\documentclass[]{spie}  
\usepackage[]{graphicx}

\usepackage{color}
\usepackage{bm}

\usepackage[english]{babel}
\usepackage{amsmath}
\usepackage{units}
\usepackage[colorlinks=true, allcolors=blue]{hyperref}
\usepackage{textcomp}
\usepackage[absolute,overlay]{textpos}
\usepackage{helvet}

\title{Anamorphic beam shaping for efficient laser homogenization: methods and high power applications} 

\author{Christoph Tillkorn\supit{a}, Andreas Heimes\supit{b}, Daniel Flamm\supit{b}, Stefan Dorer\supit{a}, Torsten Beck\supit{b}, Julian Hellstern\supit{a}, Felix Marschall\supit{b} and Christian Lingel\supit{b}
\skiplinehalf
\supit{a}TRUMPF Laser GmbH, Aichhalder Stra{\ss}e 39, 78713 Schramberg, Germany; \\
\supit{b}TRUMPF Laser- und Systemtechnik GmbH, Johann-Maus-Str.2, 71254 Ditzingen, Germany
}

\authorinfo{Further author information:\\ Christoph Tillkorn, E-Mail: christoph.tillkorn@de.trumpf.com}

\begin{document} 
  \maketitle 

\begin{abstract}
The homogenization of light is widely applied in various industrial sectors. The uniform high power processing of large areas requires a high degree of homogeneity. Sophisticated beam transformation techniques are used to optimize the illumination of standard optical diffusers such as microlens arrays and decrease the contrast of interference. Novel design techniques take advantage of a multimodal approach which is especially adapted to the characteristic properties of the laser light source. We show how anamorphic beam shaping is employed to transform the high power light source in order to meet the required level of homogeneity suited for the respective application.
\end{abstract}



\section{INTRODUCTION}
\label{sec:intro}  
\begin{textblock*}{16cm}(2.67cm,1cm) 
   \centering
  \tiny \textsf{Christoph Tillkorn, Andreas Heimes, Daniel Flamm, Stefan Dorer, Torsten Beck, Julian Hellstern, Felix Marschall, and Christian Lingel, "Anamorphic beam shaping for efficient laser homogenization: Methods and high power applications", Proc. SPIE 10518, Laser Resonators, Microresonators, and Beam Control XX, 105181I (16 February 2018); \url{https://doi.org/10.1117/12.2290154}.}
\end{textblock*}

\begin{textblock*}{5cm}(9.2cm,1.9cm) 
   \textsf{Invited Paper}
\end{textblock*}

\begin{figure}[t]
	\begin{center}
		\includegraphics[width=0.5\columnwidth]{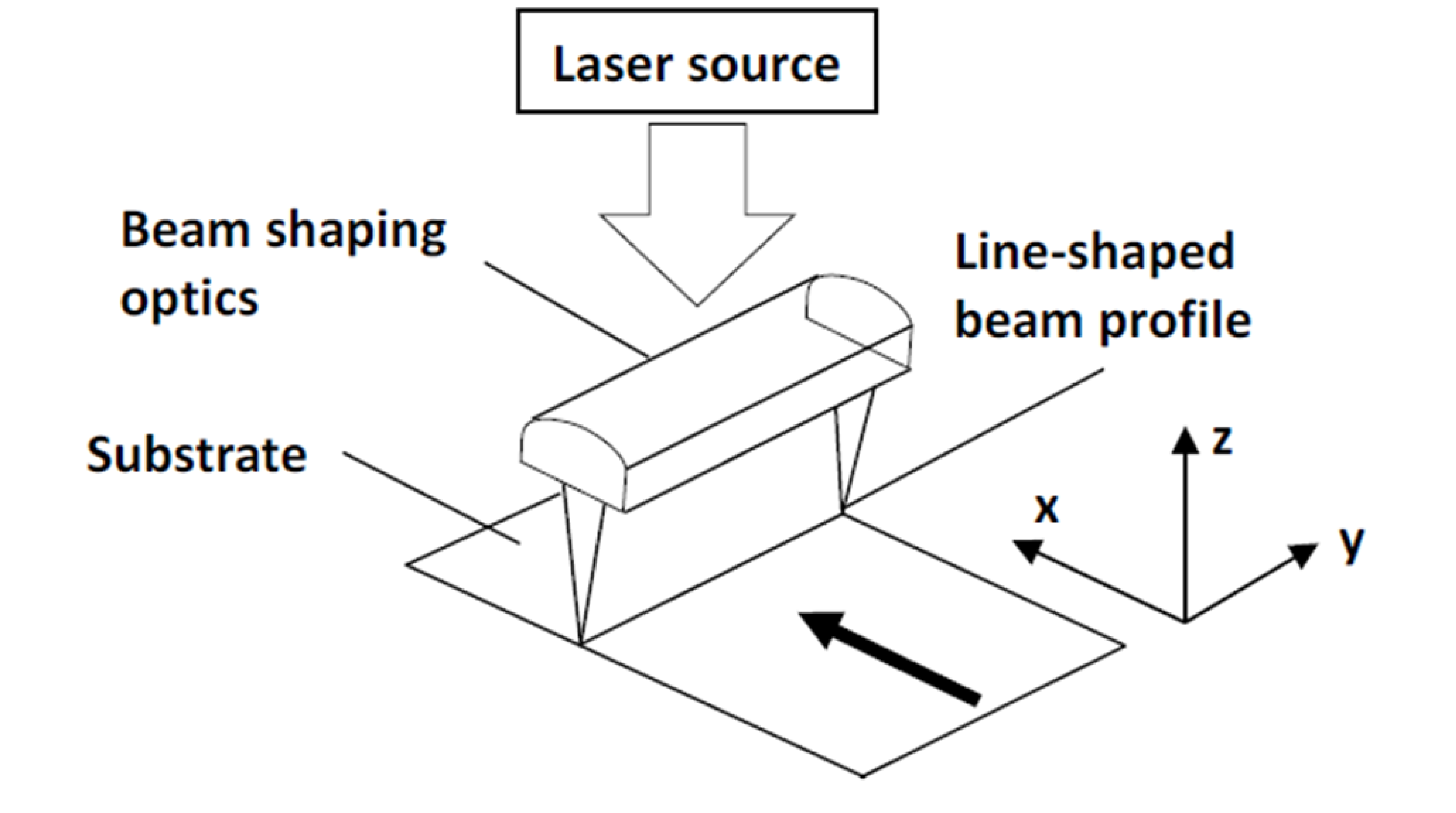}
	\end{center}
	\caption{Laser-based RTP with linear scanning and line-shaped beam profile.}
	\label{fig::setup}
\end{figure}

\begin{textblock*}{17cm}(2.25cm,25.25cm) 
   \centering \small 
   \textsf{
   © 2018 Society of Photo‑Optical Instrumentation Engineers (SPIE). One print or electronic copy may be made for personal use only. Systematic reproduction and distribution, duplication of any material in this publication for a fee or for commercial purposes, and modification of the contents of the publication are prohibited. \\
   Laser Resonators, Microresonators, and Beam Control XX, edited by Alexis V. Kudryashov, Alan H. Paxton,\\ Vladimir S. Ilchenko, Proc. of SPIE Vol. 10518, 105181I. © 2018 SPIE. \url{https://doi.org/10.1117/12.2290154}}
\end{textblock*}
The availability of robust high-power laser sources facilitates the rapid thermal processing (RTP) of large area substrates. The applications range from annealing of semiconductors \cite{Yaron1980} or the lift-off of thin layers from substrates used in the production of OLEDs \cite{Delmdahl2014} to the annealing of multi-functional coatings \cite{Tillkorn2017, Skorupa2013}. In contrast to e.g. furnace-based methods laser-based RTP has the advantage that energy can be deposited locally, i.e. to a specific area or layer. This prevents the substrate from thermal distortion or microstructural changes. In Fig.\,\ref{fig::setup} the schematic setup for laser-based RTP is shown. An anamorphic beam shaping optics generates a line-shaped beam profile which has preferably well defined dimensions in x- and y-direction. Such intensity profiles can be for instance realized by the imaging of rods, multi-aperture micro optics or by the use of diffractive optical elements. 
For high power applications on large substrates microlens arrays are a common choice of homogenization optics. These can be produced with high accuracy and offer a high efficiency at the same time providing a large angular spectrum. However, depending on the application, these optics make great demands on the laser source itself. In Fig.~\ref{fig::ill_img_hom} the basic principle of an imaging homogenizer is given. Each microlens is imaged into well defined angular spectra which overlap. In case of a totally incoherent source the intensity of the individual parts simply add without any phase information. However a certain degree of coherence leads to diffraction patterns. The latter strongly depend on the coherence properties of the source. A low degree of coherence yields less interference contrast. Either the source itself bears a high degree of incoherence or proper beam shaping techniques have to be used to achieve less spatial coherence. For either analysis, wave-optical calculations are essential \cite{Zimmermann2007,Buettner2002} that include a complete model of the light source \cite{Partanen2015}.

\par

In this paper we want to give a short summary of source models and
discuss the influence of partial coherence on the degree of homogenization achieved by standard homogenization techniques. 
In the first part we give a short introduction into the theoretical description of partially coherent light. We will introduce the basic concepts to describe spatial coherence by defining the cross spectral density (CSD) of the electric field. Furthermore we pay attention to the imaging homogenizer based on microlens arrays (MLAs) and derive a general expression for the line intensity profile. 
Using the example of different source models this result can be used to make a statement about the interference contrast which needs to be minimized depending on the specific application.
We introduce beam shaping methods to reduce the coherence of the laser source and discuss how the design of these optical elements can be optimized. Moreover we apply our model to two different partially coherent laser-light sources and compare our model with recent experimental data. In the second part of the paper we present a novel optical concept where we have applied anamorphic beam shaping to achieve a high degree of homogeneity at multi-kW power level for large area RTP.


\section{Methods}
The homogenization of laser light strongly depends on the coherence of the laser source. 
Classical optical elements based on MLAs show an interference contrast that strongly depends on these properties. 
Because of that we start this section by introducing the mathematical concept of spatial coherence and demonstrate its influence on the degree of homogenization. 

\subsection{The laser source}
This discussion is based on the so called cross-spectral-density (CSD), which describes the correlation of the electric field $\psi$ at two different source points,

\begin{figure}[t]
	\begin{center}
		\includegraphics[width=0.7\columnwidth]{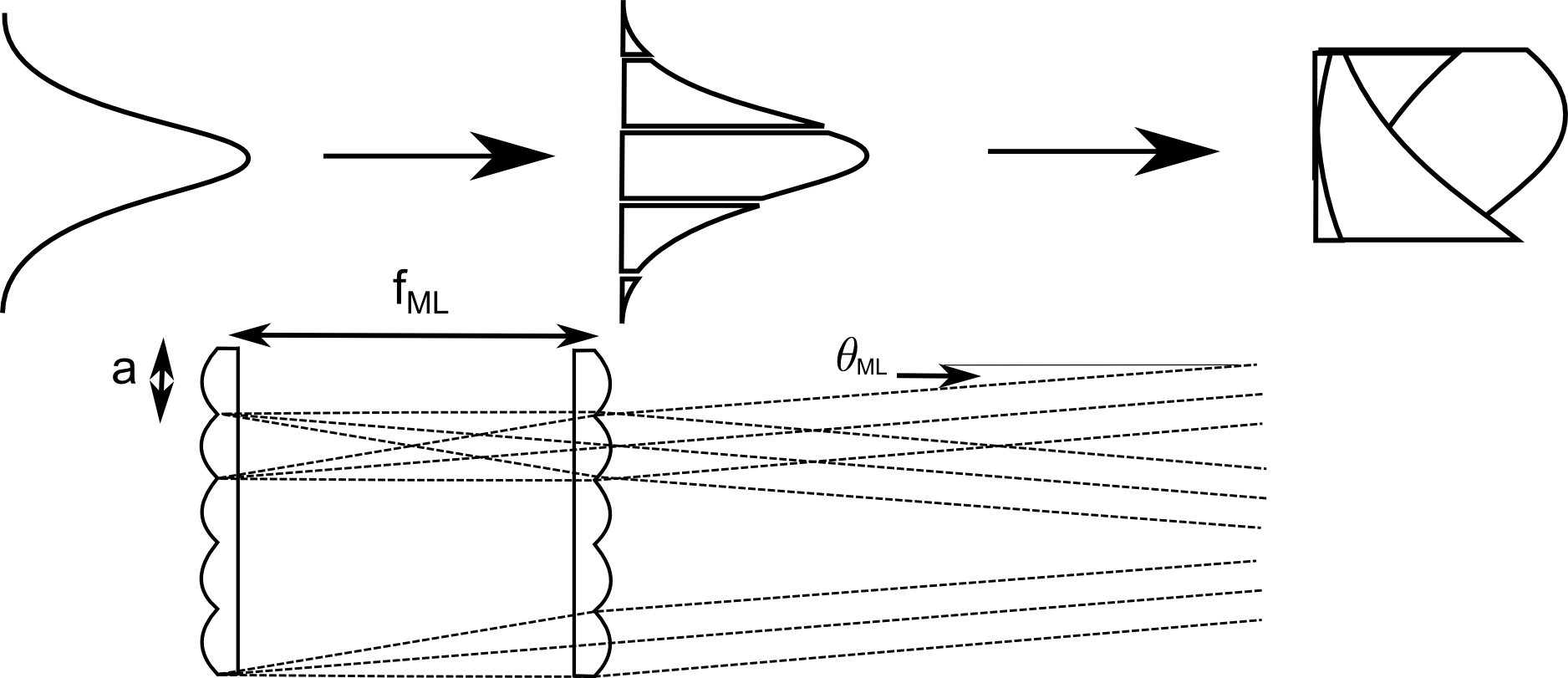}
	\end{center}
	\caption{Imaging homogenizer: Each microlens subaperture is imaged into a well defined angular spectrum. The angular spectra superimpose in far field.}
	\label{fig::ill_img_hom}
\end{figure}

\begin{eqnarray}
\label{eq::CSD}
\Gamma(\bm r_1,\bm r_2)=\langle \psi^*(\bm r_1) \psi(\bm r_2)\rangle.
\end{eqnarray}
In section \ref{section::imag_homo}, when considering the imaging homogenizer, we will see that the functional form of the CSD will directly enter the expression for the generated and homogenized angular spectrum. 
Furthermore the CSD can be derived for various source models, making this approach very universal.

Techniques for accessing the complex-valued CSD purely experimentally are often connected to the famous Young's double slit experiment evaluating contrast and position of the interference pattern caused by superimposing two or more isolated field sections of one source \cite{Born1991}. Due to the high experimental effort of the process, especially when a high spatial resolution of $\Gamma$ is demanded, several more sophisticated -- but related -- methods were developed \cite{Santarsiero2006}. Experimentally significantly simpler procedures to directly measure the CSD of a light source are based on, e.g., the determination of corresponding phase-space functions, such as the Wigner-distribution function \cite{Eppich1998}, wavefront sensing \cite{Stoklasa2014}, or holography \cite{Ferreira2013}. As we will show in the following, with a certain level of \textit{a priori} knowledge about the light source, such as the resonator design, the measurement of the beam propagation ratio $M^2$ may already be sufficient for an accurate estimation of $\Gamma$. 

Many approaches follow the modal decomposition of laser sources \cite{Forbes2016}. First of all the choice of modes can be based on the fundamental laser modes giving a physical justification. Secondly this approach is numerically feasible and, finally, there are several techniques known to directly measure the mode content of laser sources \cite{Nicholson2008, Andermahr2008, Bruening2013, Flamm2012}. In this case the CSD is given by

\begin{eqnarray}
\label{eq::CSD_modal_decomp}
\Gamma(\bm r_1,\bm r_2)=\sum_{mn}\langle c_m^* c_n \rangle\psi_m^*(\bm r_1) \psi_n(\bm r_2).
\end{eqnarray}
Here $\{\psi_n\}$ is a complete set of modes and $c_n$ are the corresponding coefficients. The set could correspond to the fundamental Gauss-modes of a laser source. Then among other things the correlation of the coefficients $c_n$ depends on the frequency distribution of the various modes. Only those possessing the same frequency have a stable intermodal phase difference \cite{Schmidt2011} and have a non vanishing expectation value $\langle c_m^* c_n \rangle$. 

In the following we will limit ourself to one dimension. Here the higher order Gauss-modes are given by $\{\psi_n(x)\}=\{{\rm HG}_n(x,w_0)\}$ with 

\begin{eqnarray}
{\rm HG}_n(x,w_0) = \left(\frac{1}{\pi w_0^2}\right)\frac{1}{(2^n n!)^{1/2}}H_n\left(\frac{x}{w_0}\right)
\exp\left(-\frac{x^2}{2w_0^2}\right), \label{eq::Gauss_Hermite}
\end{eqnarray} 
where $w_0$ is the beam radius of the lowest order mode. On the basis of the previous discussion the coefficients in Eq.~\eqref{eq::CSD_modal_decomp} are given by $\langle c_m^*c_n \rangle = \delta_{mn} |c_n|^2$. This model is suitable for, e.g., partially coherent freespace laser sources. The relation between beam quality factor and coefficients is given by  $M^2=\sum_n |c_n|^2 (2n+1)$.
A special case, i.e. $|c_n|^2 \sim (M^2-1)^n/(M^2+1)^n$, yields the 
widely used analytical representation of the CSD introduced by Schell \cite{Partanen2015,Collett1980}. In the so called Gauss-Schell model the intensity as well as the correlation function is parametrized by a gaussian envelope, i.e.

\begin{eqnarray}
I(x)= I_0 \exp\left(-\frac{2x^2}{w_R^2}\right),\qquad \mu(x_1,x_2)=\exp\left(-\frac{|x_1-x_2|^2}{2 l_c^2}\right), \label{eq::CSD_gauss_shell}
\end{eqnarray}
where $w_R=Mw_0$ is the beam radius and $l_c$ is the spatial coherence length.
The ratio of beam radius $w_R$ and coherence length $l_c$ determines the spatial coherence of the beam and thus the beam quality $M^2$, i.e. $M^2 = [{1+(w_R/l_{c})^2}]^{1/2}$ \cite{Partanen2015}. We will refer to this special case regularly in the following since the interference contrast is reduced to measurable parameters such as beam radius and coherence length. 

\begin{figure}[t]
	\begin{center}
		\includegraphics[width=1\columnwidth]{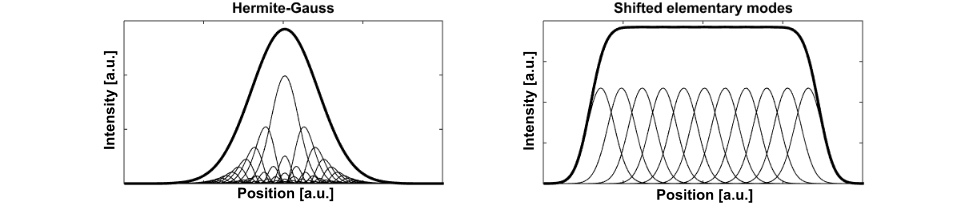}
	\end{center}
	\caption{Multimode-Models: (Left) Gauss-Schell model decomposed into a set of Hermite-Gauss modes. (Right) Top-hat in near field generated by elementary shifted Gaussian envelopes.}
	\label{fig::multimode_models}
\end{figure}

So far we only considered source models that posses similar near and far field profiles. In case of lasers source whose near and far field distributions differ such as, e.g., laser light emitted by step-index fibers, parameterizations other than Eq.~\eqref{eq::Gauss_Hermite} have to be chosen. As an example we will employ the model of elementary shifted Gauss-modes (see Fig.~\ref{fig::multimode_models}). This model is based on a set of shifted lowest order Gauss modes, i.e. $\{\psi_n(x)={\rm HG}_0(x-x_n)\}$ with $\langle c_m^*c_n\rangle =\delta_{mn} |c_n|^2$. The far field remains gaussian, whereas the near field may has an arbitrary shape. The CSD for the elementary shifted function approach is given by

\begin{eqnarray}
\Gamma_{\rm EF}(x_1,x_2) &=& \sum_n |c_n|^2 {\rm HG}^*_0(x_1-x_n,\sigma){\rm HG}_0(x_2-x_n,\sigma).
\end{eqnarray}

Here the parameter $\sigma$ controls the divergence of the beam. The divergence angle is given by $\theta_R = \lambda/\pi \sigma$.
In Fig.~\ref{fig::multimode_models} we compare the two models. Later, in this section we will also compare the Hermite-Gauss approach as well as the shifted elemantary function approach with recent experimental data. Many commercial codes allow the use of higher order Gauss modes as well as the elementary shifted modes including the choice of coefficients such as, e.g., VirtualLab Fusion \cite{Wyrowski2015}.

\subsection{The imaging homogenizer} \label{section::imag_homo}
\begin{figure}[t]
	\begin{center}
		$
		\begin{array}{c}
		\includegraphics[width=0.9\columnwidth]{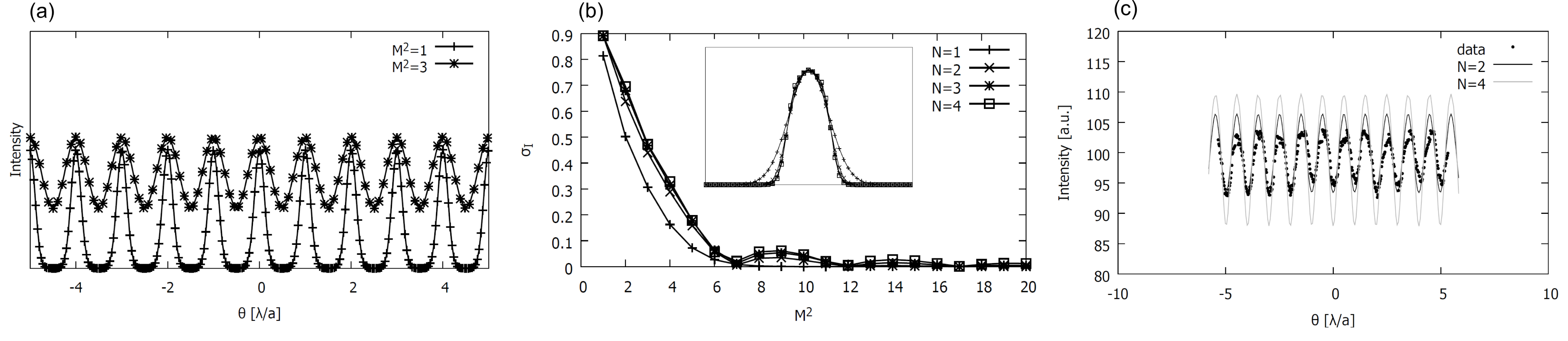}
		\end{array}
		$
	\end{center}
	\caption{(a) Intensity profile behind homogenizer. (b) Interference contrast $\sigma_I$ over beam quality $M^2$ for $\pi w_R/a = 5$ and ${\rm FN}=a^2/4\lambda f_{ML}=128$. The inset shows the beam profiles corresponding to the parameters $N=1-4$.
		(c) Fit of the model parameter to experiment. Here the parametrization defined in Eq.~\eqref{eq::coeff_HG} is used. The super gaussian factors $N=2$ and $N=4$ are shown.}
	\label{fig::ic_over_m2}
\end{figure}

The imaging homogenizer is a standard means for anamorphic beam shaping. It consists of two MLAs characterized by a certain pitch $a$ and focal length $f_{ML}$ (see Fig.~\ref{fig::ill_img_hom}). Generally the MLAs have a distance close to $f_{ML}$. This leads to the following behavior: Each ML aperture is mapped into a well defined angular spectrum, each of which is superimposed by a Fourier lens in the focal plane. In the limit of a totally incoherent illumination this leads to a homogenized angular spectrum confined by the numerical aperture of the microlenses, i.e. ${\rm NA}_{ML} =a/2f_{ML}$. For a partially coherent source the difference in optical path length leads to interference patterns with the angular period 

\begin{align}
\delta \theta = \frac{\lambda}{a}. \label{eq::periodicity}
\end{align}
In the appendix \ref{app::IntProfile} we show that an analytical expression can be derived for the generated angular spectrum behind the homogenizer. It is given by

\begin{eqnarray} 
&& I(\theta) ={\rm rect}\left(\theta \frac{2f_{ML}}{a}\right)   \sum_{lp} \exp\left(i{\frac{2\pi f_{ML}}{a} (p-l)}\theta \right) \tilde \Gamma \left(\theta - \frac{\lambda}{a}l,\theta - \frac{\lambda}{a}p\right). \label{eq::angular_spectrum}
\end{eqnarray}
Here we identify the far field transform of the CSD in Eq.~\eqref{eq::CSD},

\begin{eqnarray}
\Gamma(\theta_1,\theta_2)=\int dx_1 dx_2 \exp\left(i\frac{2\pi(\theta_1x_1-\theta_2x_2)}{\lambda}\right)\Gamma(x_1,x_2).
\end{eqnarray} 
Furthermore the window function is given by ${\rm rect}=1$ if $|x|\leq 1$ and ${\rm rect}=0$ elsewise. The additional phase factor accounts for the difference in optical path length.
Eq.~\eqref{eq::angular_spectrum} illustrates that the coherence properties of the light source, quantified by the CSD, strongly influence the interference contrast of the angular spectrum.
The far field transform of the Hermite-Gauss model and the elementary function model are given by 

\begin{eqnarray}
\tilde \Gamma_{\rm HG}(\theta_1,\theta_2) &=& \sum_n |c_n|^2 {\rm HG}_n^*(\theta_1,\theta_0){\rm HG}_n(\theta_2,\theta_0), \label{eq::CSD_HG}\\
\tilde \Gamma_{\rm EF}(\theta_1,\theta_2) &=& \sum_n |c_n|^2 \exp\left(i\frac{2\pi(\theta_1-\theta_2)x_n}{\lambda}\right){\rm HG}_0^*(\theta_1,\theta_R){\rm HG}_0(\theta_2,\theta_R), \label{eq::CSD_EF}
\end{eqnarray}
respectively. Mind that the divergency angle $\theta_0$ is given by $\theta_0 = \lambda M/\pi w_R$.
In order to get a better understanding of the relation between beam quality and the intensity distribution we refer to the Gauss-Schell-model, Eq.~\eqref{eq::CSD_gauss_shell}. For a collimated beam the angular CSD reads

\begin{eqnarray}
\tilde \Gamma(\theta_1,\theta_2)=\exp\left(-\frac{\theta_1^2}{\theta_R^2}\right)\exp\left(-\frac{\theta_2^2}{\theta_R^2}\right)
\exp\left(-\frac{\Delta\theta^2}{2\theta_c^2}\right) \label{eq::CSD_Gauss_Schell}
\end{eqnarray}
with $\theta_R= \frac{\lambda M^2}{\pi w_R} $ and $\theta_c= \theta_R/\sqrt{M^4-1}$. Plugging this equation into Eq.~\eqref{eq::angular_spectrum} yields an intensity profile which is characterized by separated gaussian envelopes as depicted in Fig.~\ref{fig::ic_over_m2}(a). The intensity follows the functional form

\begin{eqnarray}
&& I(\theta) ={\rm rect}\left(\theta \frac{2f_{ML}}{a}\right)\sum_{l} \exp\left(-\frac{2\left(\theta - \frac{\lambda}{a}l\right)^2}{\theta_R^2}\right)  \sum_p \exp\left(i\frac{2\pi f_{ML}}{a}p\theta\right)\exp\left(-\frac{p^2\lambda^2 M^4}{2a^2\theta_R^2}\right).
\label{eq::angular_spectrum_gauss_shell} 
\end{eqnarray}
In the limit 
${\theta_R}\gg {\lambda/a}$
the Gaussians envelopes in the first sum have a strong overlap. A reduction of beam quality, i.e. $M^2 \gg 1$, leads to a gradual disappearance of the phase relation in the second term. The intensity profile simply becomes homogeneous.     
The situation may differ for an angular spectrum which has a shape other than Gaussian.
Let us reconsider the parametrization of the CSD by a set of Hermite-Gauss modes, Eq.~\eqref{eq::Gauss_Hermite}. We parametrize the coefficients as follows 

\begin{eqnarray}
|c_n|^2  \sim  \exp{\left(-{n^N}/{\sigma_M^N}\right)}. \label{eq::coeff_HG}
\end{eqnarray} 
Here $\sigma_M$ controls the effective beam quality, whereas the parameter $N$ controls the steepness of the beam profile and the angular spectrum. In Fig.~\ref{fig::ic_over_m2}(b) we plot the interference contrast

\begin{eqnarray}
\sigma_I^2 = \frac{\langle I(\theta)^2\rangle -\langle I(\theta)\rangle^2}{2\langle I(\theta)\rangle^2} \label{eq::int_contr}
\end{eqnarray}
against the beam-quality $M^2$ for varying parameters $N$. The case $N=1$ corresponds to the already discussed Gauss-Schell model. With increasing beam quality factor the interference contrast is reduced. For $N>1$ the contrast recurs periodically but decreases gradually with $M^2$. This effect has been already discussed by Zimmermann et al. \cite{Zimmermann2007} and can be understood when considering Fig.~\ref{fig::ic_over_m2}(a). The intensity profile is characterized by an equidistant copy of the angular spectra of the incident light. Thus the width of each envelope is given by $\theta_R$. For fixed $\theta_c$ the divergence $\theta_R$ increases linearly with $M^2$ leading to an overlap of neighboring peaks until the modulation vanishes completely. A further increasing of $\theta_R$ leads to recurrent contrast whose amplitude strongly depends on the shape of the incident angular spectrum.   

In Fig.~\ref{fig::ic_over_m2}(c) we compare this model with a recent experimental result. The plot shows a small section for the angular spectrum behind an imaging homogenizer. The source is characterized by a given beam quality factor $M^2$ and fixed divergence angle $\theta_R$. Using the parametrization in Eq.~\eqref{eq::coeff_HG} the only free parameter that remains is the supergaussian factor $N$. Adjusting this parameter we find a good agreement with the observed interference contrast. 

\subsection{Beamtransformation}\label{section::BTS}
\begin{figure}[t]
	\begin{center}
		\includegraphics[width=0.8\columnwidth]{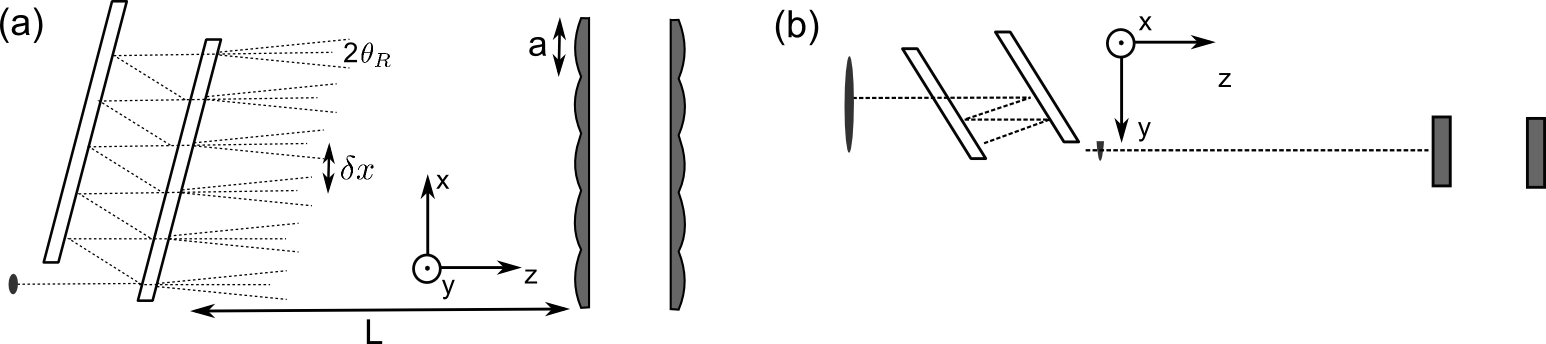}
	\end{center}
	\caption{Anamorphic beam shaping principle: (a) Top-view: in $x$-direction the beam transformation module generates $N$ separated beams, which are arranged by an equidistant spacing $\delta x$. (b) Side-view: the incoming beam is extended in $y$-direction and cut into segments whose size is defined by the exit aperture and angle of tilt. }
	\label{fig::platttrafo}
\end{figure}

Based on the previous section it seems a reasonable strategy to decrease the beam quality considerably in order to improve the beam homogenization. Many laser sources provide only a partial degree of coherence. In this case anamorphic beam shaping helps to degrade the beam quality along the homogenization axis. In \cite{Tillkorn2017} some of the authors already demonstrated the use of such a beam shaping element \cite{Clarkson96} (see Fig.~\ref{fig::platttrafo}).

\begin{figure}[t]
	\begin{center}
		\includegraphics[width=0.5\columnwidth]{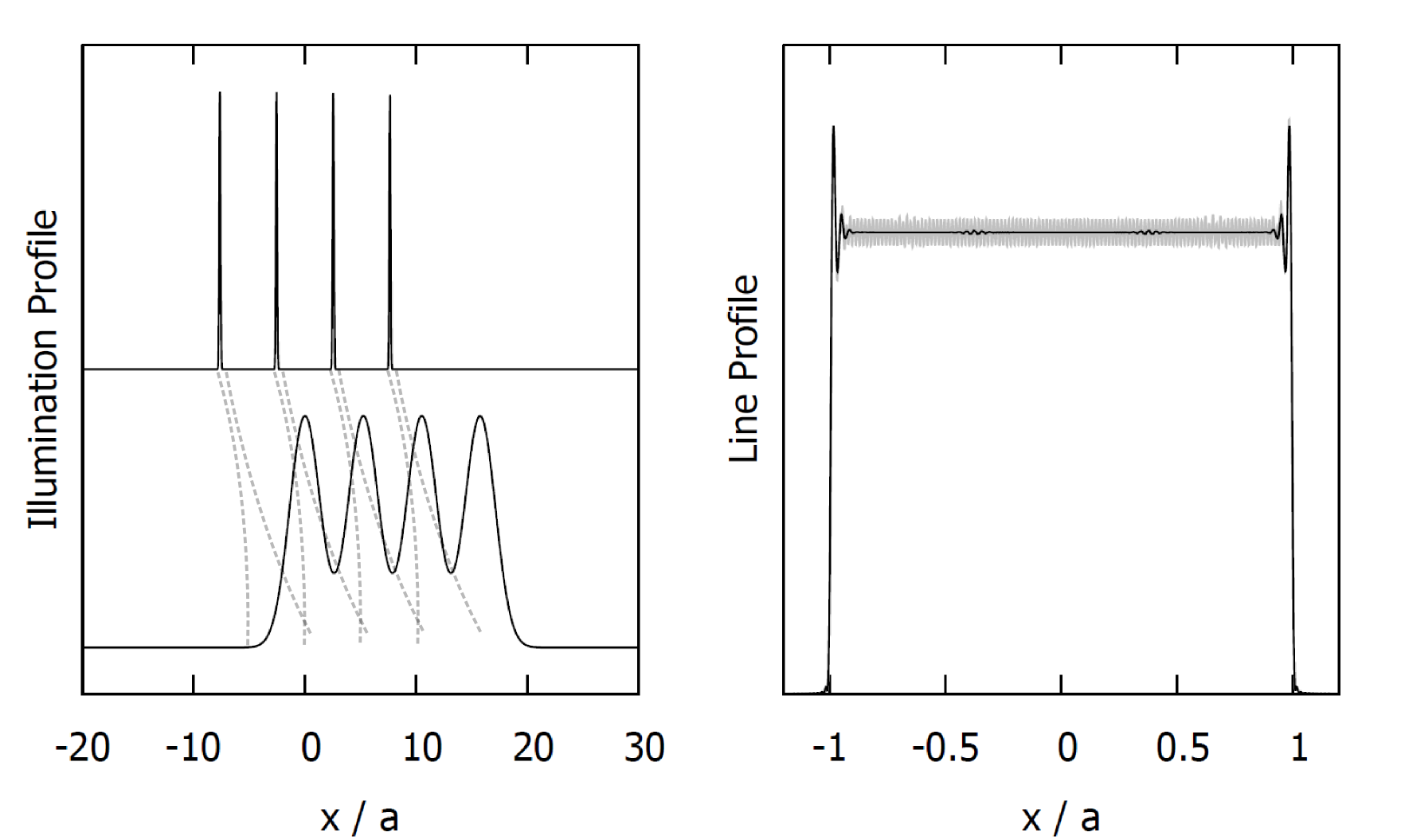}
	\end{center}
	\caption{Left: incoherent superposition of $N=4$ separated multi-mode Gauss-Schell sources with divergence $\theta_R=0.275\,\frac{a}{2f_{ML}}$ and $M^2=5$. The sources are propagated by a certain distance in order to realize a certain filling of the microlenses. Right: Homogenized laser profile. The homogenizer used has a Fresnel number of $FN=\frac{a^2}{4\lambda f_{ML}}\approx 30$. The black and grey lines correspond to a spacing of $\delta x =2a+\frac{a}{N}$ and $\delta x =2a+0.4\frac{a}{N}$.}
	\label{fig::multi_mode_bs}
\end{figure}

An elliptical beam with a well defined divergence $\theta_R$ in long axis is subjected to multiple reflections and separated into $N$ pieces by a spacing of $\delta x$. Provided that the difference in optical path length is larger than the longitudinal coherence length of the light source these beam packages are incoherent to each other. In this case the modified angular CSD reads

\begin{eqnarray}
\tilde \Gamma_{BS} (\theta_1,\theta_2)=\gamma(\Delta \theta)\cdot\tilde \Gamma(\theta_1,\theta_2) \label{eq::CSD_BS}
\end{eqnarray}
with
\begin{eqnarray}
\gamma(\Delta \theta) = \sum_l \alpha_l \exp\left(i\frac{2\pi}{\lambda} \Delta \theta x_l\right). \label{eq::CSD_BS_phase}
\end{eqnarray}
In the limit of equal weighting $\alpha_l=1$ and infinitely many randomly distributed sources the sum vanishes, unless $\Delta \theta =0$. This would correspond to a completely incoherent illumination of the homogenizer. 
In reality only a finite number of splittings is generated raising the question whether the arrangement of beams can be optimized. In this respect Eq.~\eqref{eq::CSD_BS_phase} gives a clue. E.g., for an equidistant spacing $x_l=l\delta x$ of $N$ beams the phase factor vanishes if  ${\rm mod}(\delta x,a) = a/N$ .
As an example we consider a Gauss-Schell source with $M^2=5$ which is separated into $N=4$ equidistant and equally weighted beams. In Fig.~\ref{fig::multi_mode_bs} the illumination of the homogenizer after a certain propagation distance is shown (lhs). On the rhs we present the homogenized profiles corresponding to two different spacings. We see that the interference contrast crucially depends on the separation of beams. The homogenization is optimized for ${\rm mod}(\delta x,a) = a/4$ (black curves) whereas the interference contrast is increased away from this condition (grey curves). 

\begin{figure}[t]
	\begin{center}
		\includegraphics[width=0.8\columnwidth]{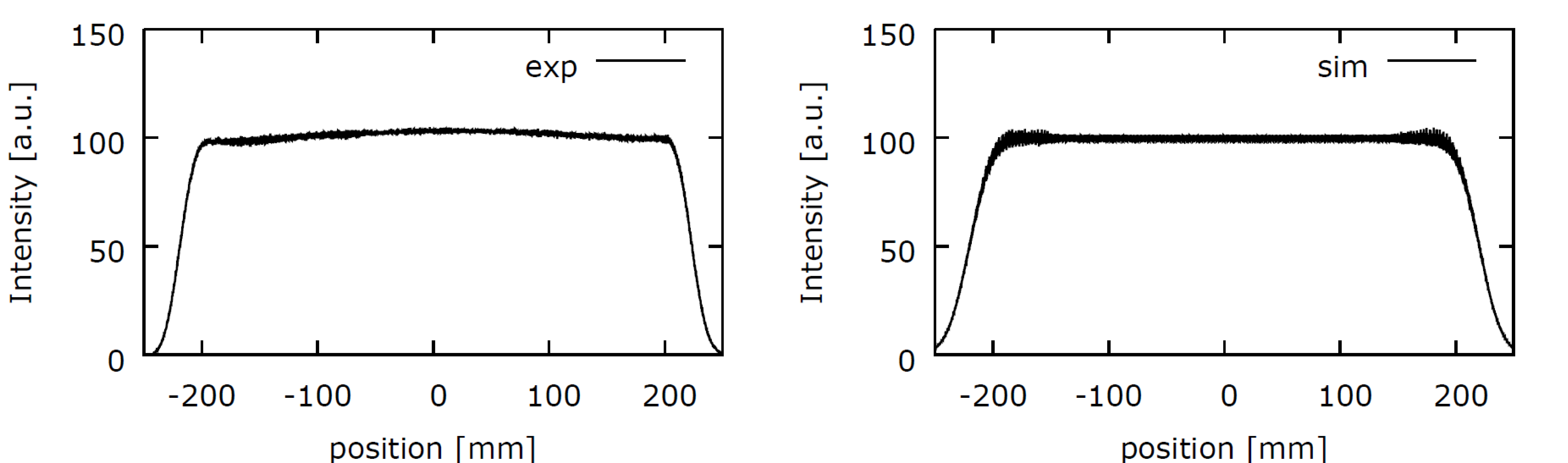}
	\end{center}
	\caption{Comparison of experimental data with simulation results based on the elementary function model.}
	\label{fig::sg_exp_sim}
\end{figure}

For the elementary shifted mode model we find that the functional form of the CSD, Eq.~\eqref{eq::CSD_EF}, is retained, i.e.

\begin{eqnarray}
\tilde \Gamma_{\rm EF}(\theta_1,\theta_2) &=& \sum_n\sum_l |c_n|^2\alpha_l \exp\left(i\frac{2\pi}{\lambda} \Delta \theta (x_n+x_l)\right)
{\rm HG}_0^*(\theta_1,\theta_R){\rm HG}_0(\theta_2,\theta_R).
\end{eqnarray}  
This is interesting because, provided that the source is described by the elementary shifted mode model, the beam transformation module simply multiplies the number of elementary modes. Therefore the model offers a useful interface between beam transformation and homogenizer. This approach is particularly interesting when many modes are involved. 
This model has been applied to a system communicated by Tillkorn et al. \cite{Tillkorn2017}. In Fig.~\ref{fig::sg_exp_sim} the elementary mode approach is compared to experimental data. To get this result, the source behind the transformer was modulated with 150 modes. The width of the elementary function source was adapted to the beam width behind the transformer.

\section{Modular Line Beam Optics} \label{section::theory_exp}
Based on the beam shaping techniques of the previous section we present a modular optical line beam concept for large area RTP which has already been introduced \cite{Tillkorn2017} and is unique with respect to its operational power level, efficiency and compactness. Due to the modular approach we are able to scale the line length in the range of meters while maintaining a low system complexity.

\subsection{Optical concept}

\begin{figure}[t]
	\begin{center}
		\includegraphics[width=0.9\columnwidth]{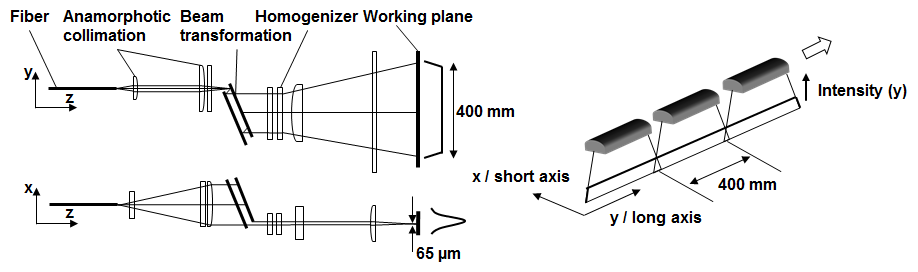}
	\end{center}
	\caption{Left: Schematic of the modular line beam optics \cite{Tillkorn2017}. Right: Stitching of individual modules.}
	\label{fig::LBO_Concept}
\end{figure}

\begin{figure}[t]
	\begin{center}
		\includegraphics[width=0.9\columnwidth]{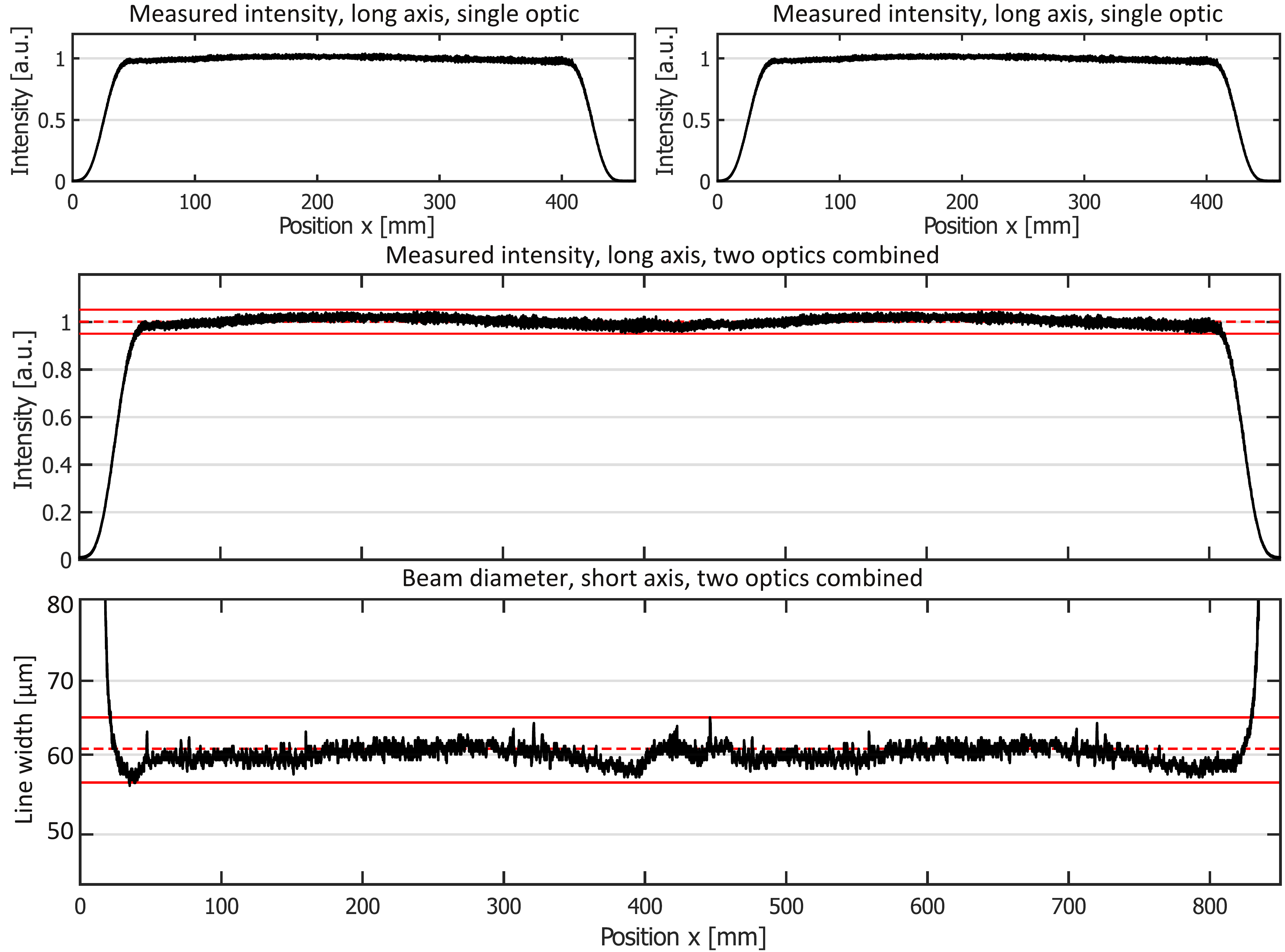}
	\end{center}
	\caption{Stitching process at 18kW input per module: (top) The single beam profiles are measured (integration in short axis). (middle) Profiles are stitched such that homogeneity requirements are met. Red lines indicate minimum and maximum intensity. (bottom) Short axis FWHM, red lines indicate minimum and maximum intensity.}
	\label{fig::stitching_measurement}
\end{figure}

The basic optical concept is given in Fig.~\ref{fig::LBO_Concept}, left, and consists of an anamorphic collimation followed by a multi-reflection beam transformation optics similar to the one proposed by Clarkson et al. \cite{Clarkson96}, an imaging homogenizer, a Fourier lens and a focusing optics. In order to provide the robustness for industrial application it relies on the established thin disk laser platform providing output powers in excess of $\unit[12]{kW}$ per unit at a wavelength of $\unit[1030]{nm}$. In the long axis a trapezoidal profile of $\unit[400]{mm}$ length with a homogeneity better than $\unit[10]{\%}$ is realized (see Fig.~\ref{fig::sg_exp_sim}). In the short axis the beam profile is quasi-gaussian having a FWHM-width of less than $\unit[65]{\mu m}$ at close to diffraction limited beam quality. Due to the compact non-telecentric design, several modules can be combined in the long axis to scale the line length by stitching of the trapezoidal beam profiles (see Fig.~\ref{fig::LBO_Concept} right, Fig.~\ref{fig::stitching_measurement} middle). Thus we have managed to overcome the limitation in processing width of approximately $\unit[1.5]{m}$ which is given by the maximum dimensions of optical components available.

The precise measurement of the line beam parameters is a key factor to meet the requirements on homogeneity for the stitched line profile. Since the tuning of the line beam parameters in practice can only be applied to an individual line beam optics, it is crucial to align every single module to specific parameters with a high degree of repeatability so that the profiles of different modules can be stitched. Therefore we have developed a CCD-based measurement setup which is able to measure the line beam profile at full power and allows to virtually stitch the profiles before installation. In Fig.~\ref{fig::stitching_measurement} a sample measurement of two line beam modules is shown (top) and virtually stitched (middle). In order to fulfill the requirements on homogeneous processing in the stitching zone it is very important to tightly control both the trapezoidal beam profile in the long axis and the uniformity of the FWHM-width of the short axis (bottom). In this case the peak-to-valley deviation of the short axis integrated intensity and the short axis FWHM-width are well below the $\unit[10]{\%}$ border so that the requirements for the specific application are met.

The key aspects determining the mechanical design are robustness and compactness and as well as thermo-mechanical stability and tightness against particle contamination to provide unrestricted suitability for 24/7 operation. Thus the  housing of the line beam module is based on a rigid aluminum plate construction having dimensions of $\unit[1000]{mm} \times \unit[400]{mm} \times \unit[300]{mm}$ at a weight of approx.~$\unit[110]{kg}$. Due to an efficiency of the beam shaping optics of larger than $\unit[90]{\%}$ thermalization of the entire system is uncritical. Moreover the compact dimensions allow for alignment with high precision alignment stages and easy integration into conveyor belts.

\begin{figure}[h!]
	\begin{center}
		\includegraphics[width=0.7\columnwidth]{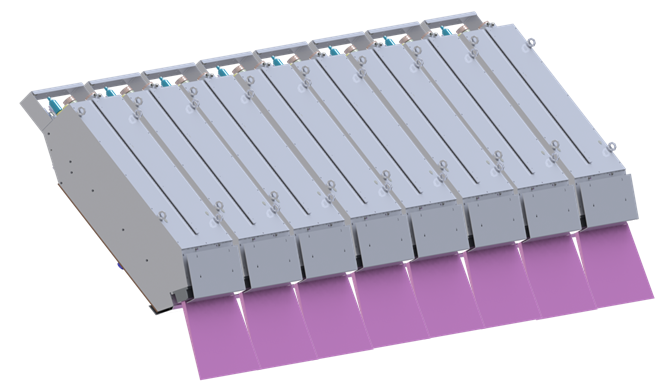}
	\end{center}
	\caption{Stitching of $\unit[8] \times \unit[400]{mm}$ line beam module.}
	\label{fig::LBO_CAD}
\end{figure}

Each module features 2 fiber ports and an electronic interface for the condition monitoring of critical parameters such as stray light and temperature. Together with our strategic partners we have realized a total line length of more than $\unit[3.2]{m}$ by combination of 8 line beam modules providing a power per unit length of more than $\unit[40]{W/mm}$ which corresponds to an input of $\unit[18]{kW}$ per individual module (see Fig.~\ref{fig::LBO_CAD}). Based on current research we can state that the current design can handle a power per unit length of more than $\unit[55]{W/mm}$ or $\unit[24]{kW}$ per module. 

\subsection{Application}

The main driver for the development of the modular line beam optics is the rapid thermal annealing of thin films deposited on jumbo-size architectural glass featuring dimensions of $\unit[3.2]{m} \times \unit[6]{m}$.

Thin coatings on glass for building applications is a highly competitive market where all the suppliers use similar technologies for coating deposition. Starting from the early eighties, the glass industry introduced magnetron cathodic sputtering as a key technology to functionalize large area glass (thermal insulation, solar reflection etc.). Contrarily to float CVD, magnetron sputtering provokes a limited thermal energy transfer to the film, leading to amorphous or poorly-crystallized material. Later, new sputtered stack designs allowed the coated glass to be tempered. The thermal heat treatment applied by the tempering process allows to improve the crystallization of metallic and some dielectric films significantly \cite{Tsang2008, Wang2005}, but remains limited to safety glass.

Compared to thermal heating by IR-lamps or furnaces, laser-based RTP allows to selectively heat the coating without affecting the glass substrate. However, in the past it could only be applied to smaller substrates.

The homogeneity achieved by the modular line beam optics now permits to process the coating homogeneously on large scale substrates at the same time fulfilling the strict optical requirements for coated glass. In addition, the large depth of focus proves to be a key to tolerate glass vertical movement during conveying under the fixed laser line. The system has also demonstrated its robustness and ability of 24/7 operation with industrial standards. In particular this implies a stable and reproducible processing speed in accordance with an inline process ($>\unit[10]{m/min}$) fulfilling efficiency requirements of the glass industry.

Due to the highly improved crystallization by laser processing, the resistivity of, e.g., thin Ag-based stacks can be decreased homogeneously up to $30\%$. This has led to the development and successful market launch of new glass products with breakthrough performance and improved energetic balance.

\section{Conclusion}
The homogenization of partially coherent light for line beam applications requires thorough wave optical modeling of the entire optical system comprising source and optical system since the interference contrast imposed by multi-aperture homogenizers critically influences the degree of homogeneity. A general description of spatial coherence is given by the cross-spectral-density (CSD). With respect to multi-mode laser sources the CSD can be represented by a set of modes making use of a priori knowledge such as the beam propagation ratio $M^2$. For the modeling of free-space lasers with quasi-gaussian near and far field the Gauss-Schell model is numerically favorable and provides very good experimental correlation. For sources with differing near and far field distributions the elementary shifted function approach is suitable. However, the number of modes must be carefully considered. Thus it is particularly interesting when many modes are involved. As a result of this we present the concept of a novel modular line beam optics which relies on the established thin disk laser technology. It allows to scale the processing width in the range of meters by stitching of individual sub-modules at a power per unit length of more than $\unit[40]{W/mm}$. As a sample application we report of the rapid thermal annealing of thin film coatings on jumbo-size architectural glass which among others has led to a new generation of low-E glass with significantly improved energy efficiency.

In the near future we plan to apply the presented simulation tools to develop optimized line beam optical systems that meet the requirements for novel laser applications.

\appendix
\section{Imaging homogenizer} \label{app::IntProfile}

The imaging homogenizer is the standard tool used for producing homogenized angular spectra.
In order to derive the angular spectrum generated behind the homogenizer we consider Fig.~\ref{fig::setup}.
The subapertures of the MLAs are imaged into well defined angular spectra, which are finally superimposed behind the homogenizer. The angular spectrum is approximately given by

\begin{eqnarray}
\Psi(\theta)=\sum_n\exp\left(i\frac{2\pi\theta na}{\lambda}\right){\rm rect}\left(\theta\frac{2 f_{ML}}{a}\right) \psi\left(\theta f_{ML}-na \right).
\end{eqnarray} 

Here $\psi(x)$ is the illumination in front of the first MLA and ${\rm rect}(\cdot)$ is the window function defined in section \ref{section::imag_homo}. The phase factor accounts for the difference in optical path between the apertures. Introducing the Fourier-transform
$\psi(x)=\int dk\,\exp(-ikx)\tilde \psi(k\lambda/2\pi)$ we obtain

\begin{eqnarray}
\Psi(\theta)= \int dk\, \exp(ik\theta f_{ML}){\rm rect}\left(\theta\frac{2 f_{ML}}{a}\right) \tilde\psi(k\lambda/2\pi)\sum_n\exp\left(-i\left(k-\frac{2\pi\theta }{\lambda}\right)na\right).
\end{eqnarray} 
The sum over the phase factor is finite ony if $\frac{k}{k_0}=\theta + \frac{\lambda}{a}m$ with $m=0,\pm 1,\pm 2,\cdots$ and therefore we obtain
\begin{eqnarray}
\Psi(\theta)&\sim & \int dk\, \exp(ik\theta f_{ML}){\rm rect}\left(\theta\frac{2 f_{ML}}{a}\right) \tilde\psi(k\lambda/2\pi)\sum_n\delta\left(\frac{k\lambda}{2\pi}-\theta + \frac{\lambda}{a}n\right)\\
&\sim & \exp\left(i\frac{2\pi}{\lambda}\theta^2 f_{ML}\right){\rm rect}\left(\theta\frac{2 f_{ML}}{a}\right) 
\sum_n \exp\left(i\frac{2\pi f_{ML}}{a}\theta n\right)\tilde\psi\left(\theta - \frac{\lambda}{a}n\right)
\end{eqnarray}
Now the generated angular spectrum behind the homogenizer is given by
\begin{eqnarray}
I(\theta) =\langle |\Psi(\theta)|^2 \rangle \sim {\rm rect}\left(\theta\frac{2 f_{ML}}{a}\right) 
\sum_m\sum_n \exp\left(i\frac{2\pi f_{ML}}{a}\theta (m-n)\right)
\left\langle\tilde\psi^*\left(\theta - \frac{\lambda}{a}n\right)\tilde\psi\left(\theta - \frac{\lambda}{a}m\right)\right\rangle.
\end{eqnarray}
This equation corresponds to Eq.~\eqref{eq::angular_spectrum} in the main text, where the expectation value at the end is transferred to the CSD.


\bibliographystyle{spiebib}   
\bibliography{Lit}


\end{document}